\documentstyle[12pt,axodraw]{article}
\newcommand{\nc}{\newcommand}
\nc{\renc}{\renewcommand}
\renc{\baselinestretch}{1.1}
\nc{\com}[1]{\ \\{\bf \# {#1}}}

\def\bort#1{ }
\newlength{\overeqskip}
\newlength{\undereqskip}
\nc{\be}[1]{\begin{equation} \mbox{$\label{#1}$}}
\nc{\bea}[1]{\begin{eqnarray} \mbox{$\label{#1}$}}
\nc{\Section}[2]{\section{\sc #2}\label{#1}\seqnoll}
\nc{\Subsection}[2]{\subsection{\sc #2}\label{#1}}
\nc{\Bibitem}[1]{\bibitem{#1}}
\nc{\Label}[1]{\label{#1}}
\nc{\eea}{\vspace{\undereqskip}\end{eqnarray}}
\nc{\ee}{\vspace{\undereqskip}\end{equation}}
\nc{\bdm}{\begin{displaymath}}
\nc{\edm}{\end{displaymath}}
\nc{\dpsty}{\displaystyle}
\nc{\bc}{\begin{center}}
\nc{\ec}{\end{center}}
\nc{\ba}{\begin{array}}
\nc{\ea}{\end{array}}
\nc{\bab}{\begin{abstract}}
\nc{\eab}{\end{abstract}}
\nc{\btab}{\begin{tabular}}
\nc{\etab}{\end{tabular}}
\nc{\bit}{\begin{itemize}}
\nc{\eit}{\end{itemize}}
\nc{\ben}{\begin{enumerate}}
\nc{\een}{\end{enumerate}}
\nc{\bfig}{\begin{figure}}
\nc{\efig}{\end{figure}}
\nc{\seqnoll}{}
\nc{\refc}[1]{\mbox{Ref.~\cite{#1}}}
\nc{\refs}[1]{\mbox{Refs.~\cite{#1}}}
\nc{\eqs}[2]{\mbox{Eqs.~(\ref{#1}) and (\ref{#2})}}
\nc{\eq}[1]{\mbox{Eq.~(\ref{#1})}}
\nc{\figs}[2]{\mbox{Figs.~\ref{#1} and \ref{#2}}}
\nc{\fig}[1]{\mbox{Fig.~\ref{#1}}}

\nc{\figcap}[1]{\begin{quote}\refstepcounter{figure}
        {\bf Figure \thefigure}: {\small #1}\end{quote}}
\nc{\tabcap}[1]{\begin{quote}\refstepcounter{table}
        {\bf Table \thetable}: {\small #1}\end{quote}}

\nc{\tag}[1]{\label{#1} \marginpar{{\footnotesize #1}}}
\nc{\mtag}[1]{\label{#1} \mbox{\marginpar{{\footnotesize #1}}}}
\nc{\etal}{\mbox{\it et al. }}
\nc{\ie}{{\rm i.e. }}
\nc{\eg}{{\it e.g. }}
\nc{\arreq}{&\!\!\!=\!\!\!&}
\nc{\arrmi}{&\!\!\!!-\!\!\!&}
\nc{\arrpl}{&\!\!\!+\!\!\!&}
\nc{\arrap}{&\!\!\!\approx\!\!\!&}
\nc{\non}{\nonumber}
\nc{\nn}{\nonumber\\}
\nc{\align}{\!\!\!\!\!\!\!\!&&}

\nc{\mat}[4]{{\left(\ba{cc} #1 & #2 \\ #3 & #4 \ea\right)}}

\def\simleq{\; \raise0.3ex\hbox{$<$\kern-0.75em
      \raise-1.1ex\hbox{$\sim$}}\; }
\def\simgeq{\; \raise0.3ex\hbox{$>$\kern-0.75em
      \raise-1.1ex\hbox{$\sim$}}\; }

\nc{\DOT}{\hspace{-0.08in}{\bf .}\hspace{0.1in}}
\nc{\Laada}{\hbox {$\sqcap$ \kern -1em $\sqcup$}}
\def\Box{\Laada}
\nc\loota{{\scriptstyle\sqcap\kern-0.55em\hbox{$\scriptstyle\sqcup$}}}
\nc\Loota{{\sqcap\kern-0.65em\hbox{$\sqcup$}}}
\nc\laada{\Loota}
\nc{\qed}{\hskip 3em \hbox{\BOX} \vskip 2ex}

\nc{\real}{{\rm I \! R}}
\nc{\Z}{{\sf Z \!\!\! Z}}
\nc{\complex}{{\rm C\!\!\! {\sf I}\,\,}}

\def\bigid{\leavevmode\hbox{\small1\kern-3.8pt\normalsize1}}
\def\id{\leavevmode\hbox{\small1\kern-3.3pt\normalsize1}}
\def\slask{\hspace{0.02em}\not\hspace{-0.27em}}
\nc{\bis}{{\prime\prime}}
\nc{\pa}{\partial}
\nc{\na}{\nabla}
\def\>{\rangle}
\def\<{\langle}

\nc{\goto}{\rightarrow}
\nc{\swap}{\leftrightarrow}
\nc{\EE}[1]{ \mbox{$\times 10^{#1}$} }
\nc{\abs}[1]{\left|#1\right|}
\nc{\at}[2]{\left.#1\right|_{#2}}
\nc{\norm}[1]{\|#1\|}
\nc{\abscut}[2]{\abs{#1}_{\scriptscriptstyle#2}}
\nc{\vek}[1]{\hbox{\boldmath$#1$}}
\nc{\integral}[2]{\int\limits_{#1}^{#2}}
\nc{\inv}[1]{\frac{1}{#1}}
\def\d#1#2{d\,^{#1}#2\,}
\nc{\dx}[1]{d\,^{#1}x}
\nc{\dy}[1]{d\,^{#1}y}
\nc{\dz}[1]{d\,^{#1}z}
\nc{\dl}[1]{\frac{d\,^{#1}l}{(2\pi)^{#1}}}
\nc{\dk}[1]{\frac{d\,^{#1}k}{(2\pi)^{#1}}}
\nc{\dq}[1]{\frac{d\,^{#1}q}{(2\pi)^{#1}}}
\nc{\dP}[1]{\frac{d\,^{#1}P}{(2\pi)^{#1}}}
\def\dbar#1#2{\frac{d\,^{#1}#2}{(2\pi)^{#1}}}

\nc{\cc}{\mbox{$c.c.$ }}
\nc{\hc}{\mbox{$h.c.$ }}
\nc{\cf}{cf.\ }
\nc{\erfc}{{\rm erfc}}
\nc{\Tr}{{\rm Tr\,}}
\nc{\tr}{{\rm tr\,}}
\nc{\pol}{{\rm pol}}
\nc{\sign}{{\rm sign}}
\nc{\bfT}{{\bf T }}

\nc{\cA}{{\cal A}}
\nc{\cB}{{\cal B}}
\nc{\cD}{{\cal D}}
\nc{\cE}{{\cal E}}
\nc{\cF}{{\cal F}}
\nc{\cG}{{\cal G}}
\nc{\cH}{{\cal H}}
\nc{\cL}{{\cal L}}
\nc{\cM}{{\cal M}}
\nc{\cO}{{\cal O}}
\nc{\cP}{{\cal P}}
\nc{\cQ}{{\cal Q}}
\nc{\cT}{{\cal T}}
\nc{\1}{\aa}
\nc{\2}{\"{a}}
\nc{\3}{\"{o}}
\nc{\4}{\AA}
\nc{\5}{\"{A}}
\nc{\6}{\"{O}}
\nc{\al}{\alpha}
\nc{\Del}{\Delta}
\nc{\e}{\epsilon}
\nc{\eps}{\epsilon}
\nc{\g}{\gamma}
\nc{\lam}{\lambda}
\nc{\om}{\omega}
\nc{\Om}{\Omega}
\nc{\ve}{\varepsilon}
\nc{\mn}{{\mu\nu}}
\nc{\ka}{\kappa}

\nc{\vp}{\varphi}
\nc{\pub}[4]{\Bibitem{#1}#2, {\sl ``#3''}, #4.}
\nc{\aap}[3]{{\it  Astron.\ Astrophys.\ }{{\bf #1} {(#2)} {#3}}}
\nc{\advp}[3]{{\it  Adv.\ in\ Phys.\ }{{\bf #1} {(#2)} {#3}}}
\nc{\annp}[3]{{\it  Ann.\ Phys.\ (N.Y.)\ }{{\bf #1} {(#2)} {#3}}}
\nc{\annraa}[3]{{\it Ann.\ Rev.\ Astron.\ Astrophys.\ }{{\bf #1} {(#2)} {#3}}}
\nc{\apl}[3]{{\it  Appl. Phys. Lett. }{{\bf #1} {(#2)} {#3}}}
\nc{\apj}[3]{{\it  Ap.\ J.\ }{{\bf #1} {(#2)} {#3}}}
\nc{\apjl}[3]{{\it  Ap.\ J.\ Lett.\ }{{\bf #1} {(#2)} {#3}}}
\nc{\app}[3]{{\it Astropart.\ Phys.\ }{{\bf #1} {(#2)} {#3}}}

\nc{\cmp}[3]{{\it  Comm.\ Math.\ Phys.\ }{{ \bf #1} {(#2)} {#3}}}
\nc{\cqg}[3]{{\it  Class.\ Quant.\ Grav.\ }{{\bf #1} {(#2)} {#3}}}
\nc{\epl}[3]{{\it  Europhys.\ Lett.\ }{{\bf #1} {(#2)} {#3}}}
\nc{\ijmp}[3]{{\it Int.\ J.\ Mod.\ Phys.\ }{{\bf #1} {(#2)} {#3}}}
\nc{\ijtp}[3]{{\it Int.\ J.\ Theor.\ Phys.\ }{{\bf #1} {(#2)} {#3}}}
\nc{\jmp}[3]{{\it  J.\ Math.\ Phys.\ }{{ \bf #1} {(#2)} {#3}}}
\nc{\jpa}[3]{{\it  J.\ Phys.\ A\ }{{\bf #1} {(#2)} {#3}}}
\nc{\jpc}[3]{{\it  J.\ Phys.\ C\ }{{\bf #1} {(#2)} {#3}}}
\nc{\jpg}[3]{{\it J.~Phys.~G:~Nucl.~Part.~Phys.~}{{\bf #1} {(#2)}{#3}}}
\nc{\jap}[3]{{\it J.\ Appl.\ Phys.\ }{{\bf #1} {(#2)} {#3}}}
\nc{\jpsj}[3]{{\it J.\ Phys.\ Soc.\ Japan\ }{{\bf #1} {(#2)} {#3}}}
\nc{\kdmfm}[3]{{\it Kong.\ Dan.\ Mat.\ Fys.\ Med.\ }{{\bf #1} {(#2)} {#3}}}
\nc{\lmp}[3]{{\it Lett.\ Math.\ Phys.\ }{{\bf #1} {(#2)} {#3}}}
\nc{\lncim}[3]{{\it  Lett.\ Nuov.\ Cim.\ }{{\bf #1} {(#2)} {#3}}}
\nc{\mpl}[3]{{\it  Mod.\ Phys.\ Lett.\ }{{\bf #1} {(#2)} {#3}}}
\nc{\naturw}[3]{{\it  Naturwiss.\ }{{\bf #1} {(#2)} {#3}}}
\nc{\ncim}[3]{{\it  Nuov.\ Cim.\ }{{\bf #1} {(#2)} {#3}}}
\nc{\np}[3]{{\it  Nucl.\ Phys.\ }{{\bf #1} {(#2)} {#3}}}
\nc{\pr}[3]{{\it Phys.\ Rev.\ }{{\bf #1} {(#2)} {#3}}}
\nc{\pra}[3]{{\it  Phys.\ Rev.\ }{{\bf A#1} {(#2)} {#3}}}
\nc{\prb}[3]{{\it  Phys.\ Rev.\ }{{\bf B#1} {(#2)} {#3}}}
\nc{\prc}[3]{{\it  Phys.\ Rev.\ }{{\bf C#1} {(#2)} {#3}}}
\nc{\prd}[3]{{\it  Phys.\ Rev.\ }{{\bf D#1} {(#2)} {#3}}}
\nc{\prl}[3]{{\it Phys.\ Rev.\ Lett.\ }{{\bf #1} {(#2)} {#3}}}
\nc{\pl}[3]{{\it  Phys.\ Lett.\ }{{\bf #1} {(#2)} {#3}}}
\nc{\prep}[3]{{\it Phys.\ Rep.\ }{{\bf #1} {(#2)} {#3}}}
\nc{\prsl}[3]{{\it Proc.\ R.\ Soc.\ London\ }{{\bf #1} {(#2)} {#3}}}
\nc{\ptp}[3]{{\it  Prog.\ Theor.\ Phys.\ }{{\bf #1} {(#2)} {#3}}}
\nc{\ptps}[3]{{\it  Prog.\ Theor.\ Phys.\ suppl.\ }{{\bf #1} {(#2)} {#3}}}
\nc{\physa}[3]{{\it  Physica\ A\ }{{\bf #1} {(#2)} {#3}}}
\nc{\physb}[3]{{\it  Physica\ B\ }{{\bf #1} {(#2)} {#3}}}
\nc{\phys}[3]{{\it Physica\ }{{\bf #1} {(#2)} {#3}}}
\nc{\rmp}[3]{{\it  Rev.\ Mod.\ Phys.\ }{{\bf #1} {(#2)} {#3}}}
\nc{\rpp}[3]{{\it Rep.\ Prog.\ Phys.\ }{{\bf #1} {(#2)} {#3}}}
\nc{\sjnp}[3]{{\it Sov.\ J.\ Nucl.\ Phys.\ }{{\bf #1} {(#2)} {#3}}}
\nc{\spjetp}[3]{{\it Sov.\ Phys.\ JETP\ }{{\bf #1} {(#2)} {#3}}}
\nc{\yf}[3]{{\it Yad.\ Fiz.\ }{{\bf #1} {(#2)} {#3}}}
\nc{\zetp}[3]{{\it Zh.\ Eksp.\ Teor.\ Fiz.\ }{{\bf #1} {(#2)} {#3}}}
\nc{\zp}[3]{{\it Z.\ Phys.\ }{{\bf #1} {(#2)} {#3}}}
\nc{\zpc}[3]{{\it Z.\ Phys.\ C\ }{{\bf #1} {(#2)} {#3}}}
\nc{\ibid}[3]{{\sl ibid.\ }{{\bf #1} {#2} {#3}}}
\def\vp{\vec{p}}

\def\Psibar{\overline{\Psi}}

\def\qh{\hat{q}}
\def\Qh{\hat{Q}}

\renewcommand{\baselinestretch}{1.25}
\begin{document}
\thispagestyle{empty}
\vfill
\eject
\begin{flushright} SUITP-98-11 \\
SUNY-NTG-98-54 \\
September, 1998
\end{flushright}
\vskip 1.5cm
\begin{center}
{\Large\bf A Simple Derivation of the \\[2mm]
Hard Thermal Loop Effective Action }
\vskip 1.2 cm
{\bf Per Elmfors$^1$} and {\bf T. H. Hansson\footnote{Supported
by the Swedish Natural Science
Research Council.} }  \\
Institute of Theoretical Physics\\
University of Stockholm \\
Box 6730, S-113 85 Stockholm, Sweden \\
elmfors@physto.se,
hansson@physto.se\\
\vskip 3mm
{\bf Ismail Zahed\footnote{Supported
by the US Department of Energy under Grant  No. DE-FG02-88ER40388.}
 }\\
Department of Physics and Astronomy\\
SUNY at Stony Brook \\
Stony Brook, New York, 11794, USA \\
zahed@nuclear.physics.sunysb.edu\\
\end{center}

\vskip 1cm
\centerline{\bf ABSTRACT}
\vskip .5cm
\noindent
We use the  background field method along with a special
gauge condition, to derive the hard thermal loop effective action in
a simple manner. The new point in the paper is to relate the
effective action explicitly to the $S$-matrix from the onset.
\vfill
\eject
\newpage
\setcounter{page}{1}
\Section{s:intro}{Introduction}
It is by now well established that the resummation of  the so called hard
thermal loops (HTLs) is a necessary part of any perturbative
scheme to finite temperature QCD. Since the original work by
Pisarski\cite{pisa1}, Braaten and Pisarski\cite{braa1} and Frenkel and
Taylor\cite{frenk1},  there have been several papers deriving the  HTL
amplitudes using various techniques, including Chern-Simons
eikonal\cite{nair1}, color transport theory\cite{blaiz1}, and Wong
equations\cite{KellyLLM94}. It has also been shown that all the HTL amplitudes
can be derived from a simple gauge-invariant effective action that
incorporates the Ward-identities originally derived by Braaten and
Pisarski\cite{braa1} and Frenkel and  Taylor\cite{frenk1}.
This effective action has been given in several forms
first by Taylor and Wong\cite{tayl1}  who gave an expression involving
string operators, and  by Braaten and Pisarski, who found the following
particularly elegant  expression ($SU(N)$ Yang-Mills theory)
\be{HTLEA}
\Gamma_{\rm HTL}[A_\mu] =  \frac{g^2NT^2}{6} \int \d4x
        \int \frac{\d2\qh}{4\pi} \,
        \tr F^{\mu\sigma} \frac{\Qh_\sigma \Qh_\lambda }
        {\left[\Qh\cdot D(A)\right]^2} F^\lambda_{~~\mu}~~,
\ee
where
$D(A)_\mu =  \partial_\mu +gA_\mu$ with $A$ in the adjoint representation,
and $\Qh$ a 4-vector of
the form $\Qh=(1,\hat q)$, so $\Qh^2=0$.%
\footnote{
We shall always use Minkowski metric, so an Euclidean 4-vector is of
the form $(iq_0,\vec q)$.}
This form was guessed by Braaten and Pisarski based on general properties
of perturbation theory, and indeed was shown to reproduce the HTL Ward
identities in the latter.
At the same time Frenkel and Taylor derived essentially the same
result by proving that the action satisfies certain conditions that are
restrictive enough to have a unique solution\cite{frenk2}.
Again their analysis was based on an analysis of the explicit (one-loop)
$n$-gluon amplitudes in the HTL approximation.

The purpose of this paper is to give a simple derivation of \eq{HTLEA},
using the background field method and the specific gauge choice
\be{gcond}
        \Qh_\mu A^{\mu} = 0~~.
\ee
It is crucial in the derivation to show that it is possible to use a $\Qh$
dependent gauge choice. Using this gauge we can  solve for the gauge
potential,
\be{Asolve}
        A^\mu(\Qh) =  \frac 1 {\Qh\cdot D(A)} \Qh_\nu F^{\nu\mu}~~,
\ee
where the parametric dependence on the gauge condition is shown
explicitly. Substituting in \eq{HTLEA}, we obtain
\be{sa}
        \Gamma_{\rm HTL}[A_\mu] =  \frac{g^2NT^2}{6}\int\d4x
        \int \frac {\d2\qh}{4\pi}
        \, \tr A(\Qh;x)^2~~,
\ee
as was first shown by Frenkel and Taylor\cite{frenk2}, who also stressed
that nothing is gained in simplicity by rewriting
$\Gamma_{\rm HTL}$ in this gauge, since the full non-local and
non-linear structure is hidden in the complicated parametric 
$\Qh$-dependence of $A^\mu(\Qh)$.

It is also not clear that the gauge in \eq{gcond} would be useful in
trying to derive the effective action, in spite of the simple form of
\eq{sa}, since it involves an integral over the parameter $\Qh$ that
enters in the gauge condition. Although, using the background field
method, the effective action by construction is invariant under
background gauge transformations, it is by no means obvious that
it can be expressed as an integral over $\Qh$ of a
gauge invariant object, as in \eq{HTLEA}.
The key ingredient in the derivation to follow in this paper
is to find such a formulation by relating   the effective action to a
gauge invariant quantity, namely the $S$-matrix. The $\qh$ integral
is naturally interpreted as an  integral over the
(on-shell) momenta $Q$ of the particles in the heat bath.

Many of the results used and/or derived in this paper have
already appeared in the literature. For example, the connection between
$\Gamma_{\rm HTL}$ and forward scattering amplitudes was already
stressed by Frenkel and Taylor, and the gauge condition
\eq{gcond} was discussed in \cite{frenk2}
and used to simplify the derivation of the
effective action in \cite{FrenkelGT95}.
We believe, however, that our approach, where the
starting point is a direct relation between the effective action and
the $S$-matrix, is novel and provides a very simple and physical
way to derive \eq{HTLEA}.

When facing the problem of how to express the HTL effective action in
terms of a gauge invariant object, it is important to understand its
physical significance beyond the formal definition of being the
generating functional of the O($T^{2}$) parts of the proper $n$-point
functions. For static fields, $\Gamma_{{\rm HTL}}$ is nothing but the
O($T^{2}$) contribution to the (negative) pressure of thermal
particles interacting with the field. For time-dependent
configurations the current is the natural object to relate to the
$S$-matrix, as was stressed in this context by Jackiw and Nair
\cite{JackiwN93}. The action $\Gamma_{\rm HTL}$ is then obtained by
integration with respect to the gauge field.
In the next section we shall first consider the static case and then the
time-dependent one. It is worth remembering that although the final formulae
derived in these cases are  the same, the physical interpretations are rather
different.

The detailed derivations in the next  section will be given for
adjoint scalars rather than gauge particles. From the presentation it
will be quite obvious that the only thing that matters for the HTL
effective action is the number of physical degrees of freedom, and their charge
and statistics. For those readers who are convinced  that this is the
case, the result for a complex adjoint scalar can immediately be taken
over to the case of YM theory since the gluon also has two physical
degrees of freedom. For those who want a formal proof, this is
provided in Appendix A using HTL power counting arguments. There, we
also prove that the HTL effective action, as calculated using
background field technique, is independent of the quantum gauge fixing
parameter. The extension of our method to include fermions is straightforward
and can be found in Appendix B.

\Section{s:F}{The free energy and the effective action}
As mentioned above, we shall consider a heat bath  of charged scalars
in the background of a non-abelian field $A_\mu$.
By construction, the one-loop background field effective action for
static field
configurations is nothing but the free energy of a gas of scalar
particles interacting with the background. The free energy may
in turn be directly related to the $S$-matrix \cite{DashenMB69}.
Hence,
\be{FdEdS}
        \Gamma^{\rm stat}_{\rm HTL}=\inv\beta\Tr_\beta\ln[-\Box]
        = F =
        F_0-\inv\beta\int_0^\infty dE\,e^{-\beta E}\inv{4\pi i}
        \tr\left(S^\dagger\frac{\pa S}{\pa E}-
        S\frac{\pa S^\dagger}{\pa E}\right)_C
        ~~,
\ee
where the trace is over all connected diagrams in the notation of
\cite{DashenMB69}. For particles that do not
interact mutually, but only with the external field, the sum over
multiparticle states can be performed and the free energy can be related to
the one-particle density of states
\be{F1p}
        F=\inv\beta\int_0^\infty
        dE\,\ln(1-e^{-\beta E})(\rho_0+\Delta\rho(E))~~,
\ee
where the shift of the density of states is related to the one-particle
$S$-matrix by
\be{dens}
        \Delta\rho(E)=\inv{4\pi i}\tr\at{\left(S^\dagger\frac{\pa S}{\pa E}-
        S\frac{\pa S^\dagger}{\pa E}\right)}{\rm 1-part}~~.
\ee
Since the $S$-matrix is gauge invariant for each physical momentum state
we can use different gauge choices for different momenta and thus the choice
in \eq{gcond} is allowed.
Then, only the first two terms in
\be{Sdiag}
\begin{picture}(40,0)(100,0)
        \SetScale{2.2}\setlength{\unitlength}{0.7661mm} 
        \Text(-3,-5)[r]{$S=$}
        \Line(5,-5)(25,-5)
        \Text(30,-5)[]{$+$}
        \Line(35,-5)(55,-5)
        \Photon(40,5)(45,-5){1}{4}
        \Photon(50,5)(45,-5){1}{4}
        \Text(60,-5)[]{$+$}
        \Line(65,-5)(85,-5)
        \Photon(75,5)(75,-5){1}{4}
        \Text(90,-5)[]{$+$}
        \Line(95,-5)(115,-5)
        \Photon(101,5)(101,-5){1}{4}
        \Photon(109,5)(109,-5){1}{4}
        \Text(120,-5)[l]{$+\ldots$}
\end{picture}
\ee
contribute. All other diagrams are either zero because of the gauge choice
or suppressed at high temperature. The counting here is very much the same
as in Appendix~\ref{a:A}.
A direct expansion of $S=\id+iT$  in a static background gives
\be{qSq}
        \<q|S(E)|q\>=1+\frac{i g^2 N}{E_q V} 2\pi\delta(E_q-E)\int \d3x
        (A^a_\mu(\Qh;x))^2~~,
\ee
using the normalization $\<q|q'\>=\frac{(2\pi)^3}{V}\delta^{(3)}(q-q')$ and
$\tr=\frac{V}{(2\pi)^3}\int\d3q$. From \eq{dens}
we obtain
\be{rho}
        \Delta\rho(E)=\frac{g^2N}{2\pi^2}\int \d3x 
        \int \frac {\d2\qh}{4\pi} (A^a_\mu(\Qh;x))^2~~.
\ee
The energy integral in \eq{F1p} is then trivial and
we arrive at
\be{Ffin}
        F=
        F_0-\frac{g^2NT^2}{12}\int \d3x 
        \int \frac {\d2\qh}{4\pi} (A^a_\mu(\Qh;x))^2~~,
\ee
which agrees with \eq{sa} when we use the relation $\Gamma=-\int_0^{-i\beta}
dx_0 F$.

The calculation above can be generalized to time dependent background fields
but, as already mentioned, the physical interpretation is different. The free
energy is an
equilibrium concept and we shall instead start from the expectation value of
the current in a background field to derive an effective action.
The current is given by \cite{JackiwN93,YZ96}
\be{current}
        j(t,x)=\inv Z\tr [e^{-\beta H} U^\dagger(t,-\infty)
        \hat \jmath(-\infty,x) U(t,-\infty)]=\inv{Z}
        \tr [e^{-\beta H}S^\dagger
        \frac{-i\delta}{\delta A(t,x)}S]~~,
\ee
with the $S$-matrix in the interaction picture.
There are two pieces in the current when written in terms of the 
$T$-matrix $S=\id +iT$
\be{T}
        j(t,x)=\inv Z\tr  e^{-\beta H}[\frac{\delta T}{\delta A(t,x)}
        -iT^\dagger \frac{\delta T}{\delta A(t,x)}]~~,
\ee
The first term is a total derivative of
an action (which we call the effective action
for time-ordered $n$-point functions).
To one-loop order the second term is imaginary and has support only when
the external field is on the light-cone. This is the term that makes up for
the difference between time-ordered and retarded $n$-point
functions.

Looking only for the real part we can
integrate the first piece in \eq{T} with respect to $A$.
We obtain
\bea{eGamma}
        e^{i\Gamma[A]}&=&\inv Z\tr[e^{-\beta H} S]=
        \inv Z\exp\biggl[V\int\dbar3q \ln\biggl(\sum_{n_q}
        e^{-\beta E_q n_q}\<n_q| S|n_q\> \biggr)\biggr]~~.
\eea
where $|n_q\>$ is a state with $n_q$ particles of momentum $q$.
Since the particles do not interact with each other the expectation value of
the $S$-matrix factorizes
$\<n_q|S|n_q\>=(\<q|S|q\>)^{n_q}=(1+i\<q|T|q\>)^{n_q}$. The sum over $n_q$
and the internal group indices can be performed easily and we find
\be{logG}
        i\Gamma[A]=-\ln Z+VN\int\dbar3q \ln
        \biggl(\inv{1-e^{-\beta E_q}(1+i\<q|T|q\>)}\biggr)~~.
\ee
Since the expectation value of the $T$-matrix is gauge invariant we
can again
choose the gauge $Q_\mu A^\mu=0$. Only the $g^2A^2$ piece in the
interaction
enters and in fact only a single such insertion  since multiple insertions
are again suppressed by powers of $1/T$. An expansion in $\<q|T|q\>$ gives
\be{linT}
        i\Gamma[A]=
        iVN\int\dbar3q \inv{e^{\beta E_q}-1} \<q|T|q\>~~.
\ee
Using \eq{qSq} for non-static background fields one finds
\be{qTq}
        \<q|T|q\>
        =\inv{E_q V}\int \d4x A^2(\Qh;x)~~,
\ee
After substituting into \eq{linT} and performing the $q$-integration, we have
\be{GammaAA}
        \Gamma_{\rm HTL}[A]=\frac{g^2 N T^2}{12}
	\int \d4x (A^{a}_{\mu})^2(\Qh;x)~~,
\ee
which again agrees with \eq{sa}. This completes our derivation of the
HTL effective action for general field configurations.\footnote{%
As usual, the form \eq{HTLEA} is appropriate only off the light cone.}
Finally, we want to stress the simplicity of the arguments leading
from \eq{FdEdS} to \eq{Ffin} and from \eq{current} to \eq{GammaAA}.
Basically all steps are written out, and the only thing that requires
some care is to get the various normalizations of the $S$-matrix
elements right. This should be contrasted with the rather involved
chain of arguments that have appeared in previous derivations of the
HTL effective action. 
This paper was concerned with a new derivation of known results, but one
could also try to use our methods to calculate subleading terms by including
interactions between the fast thermal particles. This would amount to
include contributions from the 2-body, 3-body etc $S$-matrix. Whether or not
this would lead to physically intersting approximations remains to be seen.
\vskip 2mm \noindent
{\bf Acknowledgement}: We thank Rob Pisarski for commenting on an early version of this
paper. I.Z. also thanks Rob Pisarski and Hidenaga Yamagishi for 
discussions.

\appendix
\Section{a:A}{Appendix}

We shall now formally show that the leading high
$T$ contribution to the one-loop
effective action for gluons is the same as for a single complex
scalar field in the adjoint representation.  Again we stress that
this is rather straightforward from the calculation in Section~\ref{s:F}.
We start from the Euclidean formulation of
finite temperature field theory, and, in the notation of \cite{hans1},
write the one-loop finite $T$, gauge
invariant background field effective action as
\be{QCD}
i\Gamma[A,\alpha] =  - {\rm Tr}_\beta\ln\left[ -\Box \right] +
   \inv2    {\rm Tr}_\beta\ln\left[-\Box g_{\mu\nu} -2g
   F_{\mu\nu}  -\left(\frac 1 \alpha -1\right)
        D_\mu D_\nu \right] \ \ \ \ \ ,
\ee
where $\Box=D_\mu D^\mu$,
$gF_{\mu\nu}=[D_\mu , D_\nu]$,
$\alpha$ is the (covariant background field)
gauge-fixing parameter, and the trace is over  color and Lorentz
indices as well as over spacetime.
The background field
effective action is by
construction gauge invariant, with respect to the background field
$A^\mu$, but to prove physical gauge invariance one must also establish
$\alpha$-independence. We first consider the background
field Feynman gauge, $\alpha = 1$,
expand the gluon trace in powers of
$F^{\mu\nu}/\Box$, and combine the leading term with that from the
ghosts to get,
\be{iGamma}
i\Gamma[A,1] =   {\rm Tr}_\beta\ln\left[ -\Box \right] +
   {\rm Tr}_\beta\left[ \frac 1 \Box 2 F^{\mu\nu} \right] -
   \inv{2} {\rm Tr}_\beta\left[ \frac 1 \Box 2 F^{\mu\rho}  \frac 1 \Box 2
F^{\rho\nu}     \right]+ \ldots \ \ \ ,
\ee
where $\Box=\pa^2+g\{\pa_\mu,A_\mu\} +g^2 A^2$.
Now recall the rules for power counting in HTL.
Each gluon propagator  contributes a term $\sim
1/q^2$, so naively it would be expected to give a suppression $\sim
T^{-2}$ to the amplitude, but as stressed by
Braaten and Pisarski, this is  not
correct. The leading contribution arises when the loop momentum is
large but the propagators also almost on shell. Performing the $q_0$
integration by closing the contour puts one propagator on shell, \ie
 $q_\mu = q\Qh_\mu $, contributing a factor $1/2q$ to the $dq$
integration, while all  the other propagators contribute with
denominators $(\Qh +p_i)^2=2q \Qh\cdot p_i + p_i^2 \approx 2q \Qh\cdot
p_i$. In this approximation the $dq$ integration factorizes and
immediately gives
the $T$ behavior of the graph by power counting.
The leading contribution to a diagram with $m$ 3-gluon vertices, $n$ 4-gluon
vertices, and $l$ insertions of $F^{\mu\nu}$ is $\sim
T^4T^mT^{-2-(m+n+l-1)}=T^{3-n-l}$ where the contributions are from the
integration measure, the momentum dependence of the 3-gluon vertices
and the propagators respectively. Note that the terms
corresponding to $l=1$ is   $\sim F^{\mu\rho}$ and thus zero because
of the Lorentz trace, and since the $l=2$ term is already at most $\sim T$,
the HTL action comes entirely from the graphs with no $ F^{\mu\nu}$
insertions.  This conclusion is independent of the gauge choice for
the background field. Thus, the YM effective action $\Tr_\beta\ln[-\Box]$
can be calculated using charged scalar inside the loop and multiplying with
the appropriate group factors.

Finally we show that the $\alpha$-dependence in \eq{QCD} is suppressed by
powers of $1/T$. Following \cite{vilk,hans1} we write
\be{gv}
\frac{\partial\Gamma}{\partial\alpha} = \inv2 {\rm
Tr}_\beta\left(\frac 1 \Box   D_\mu {\cal E}_\mu\frac 1 \Box \right)
 -\inv2 {\rm Tr}_\beta\left(\frac 1 \Box {\cal E}_\mu
G_{\mu\nu}{\cal E}_\nu\frac 1 \Box\right)~~,
\ee
where $ {\cal E}_\mu = [D_\nu , F_{\mu\nu}]$ and $G$ the full covariant
gluon propagator satisfying
\be{ge}
        \left(\Box g_{\mu\lambda} + 2F_{\mu\lambda} + \frac {1-\alpha} \alpha
        D_\mu D_\lambda \right) G_{\lambda\nu}(x,y) =
        g_{\mu\nu}\delta^4(x-y)~~.
\ee
It is easy to see that, since ${\cal E}_\mu$ is independent of the loop
momentum and the presence of a double pole in the first term
in \eq{gv} gives an extra power of $T$ compared to diagrams with
only single poles, there is no $T^2$ contribution.
In the second term we can expand $G$ like
\be{Gexp}
        G=\sum_n\left(\inv{\Box}(2F_{\mu\nu}+\frac {1-\alpha}
        \alpha D_\mu D_\nu)\right)^n \inv{\Box}~~.
\ee
Now we have potentially dangerous terms with powers of $\inv{\Box}D_\mu
D_\nu$ which naively go like $T^n$. However, the $D$ factors can always be
commuted around so that a contraction is possible. For example
\be{contract}
        \inv{\Box}D_\alpha D_\beta\inv{\Box} D_\beta D_\gamma=
        \inv{\Box}D_\alpha D_\gamma+
        \inv{\Box}D_\alpha
        D_\beta\inv{\Box}[D_\beta,\Box]\inv{\Box}D_\gamma~~,
\ee
where the second term goes like $T$ since
$[D_\beta,\Box]=g (D_\mu F_{\beta\mu}+F_{\beta\mu}D_\mu)\sim T$.
Each contraction lowers the naive power by one factor of $T$
and therefore
the dangerous powers can be eliminated. Terms with factors of $F$ are of
course subleading according to  the same power counting.

\Section{a:fermions}{Appendix}
It is not hard to generalize the above argument to include
fermions by adding a background of fermionic
sources. In the Feynman gauge the total effective action is
\be{fermEA}
        i\Gamma[A,\Psibar,\Psi]=- {\rm Tr}_\beta\ln\left[ -\Box \right] +
        \inv 2 {\rm STr}_\beta\ln\left[
        \begin{array}{cc} -\Box g_{\mu\nu} -2gF_{\mu\nu} &
        ig\Psibar\gamma_\nu \\
        ig\gamma_\mu\Psi & i\slask  D \end{array} \right]~~,
\ee
where the second trace now is a supertrace over both gauge bosons and
fermions. This is related to the $S$-matrix in the same way as \eq{FdEdS}
and therefore we can use the same $Q$-dependent gauge as in \eq{gcond}.
It can be rewritten in terms of an ordinary trace as \cite{DeWitt87}
\bea{fEAexp}
        i\Gamma[A,\Psibar,\Psi]
        &=&\inv 2 {\rm Tr}_\beta\ln\left[-\Box g_{\mu\nu} -2gF_{\mu\nu}\right]
        -\inv 2 \Tr_\beta\ln\left[i\slask D\right]\nonumber\\
        &&-\inv 2 \Tr_\beta\ln\left[1-\frac{g^2}{i\slask D}
        \gamma_\mu \Psi\left(\inv{\Box +2gF}\right)_{\mu\nu}\Psibar\gamma_\nu
        \right]~.
\eea
The second term is the contribution to the gauge boson effective action
from dynamical fermions which is simply equal to
$-\inv4\Tr_\beta\ln[-\Box]$. The only difference with the term
evaluated earlier is the statistics of the hard particles and 
that the group trace is in the fundamental
representation and gives a factor $N_f/2$.
The last term in \eq{fEAexp} gives  the
effective action for the fermionic background fields. It can also be
analysed with the methods described above. After expanding in powers of
$\Psibar\Psi$ and in powers of $F$
it is only the zeroth order term
\be{fterm}
        N \,\tr_\beta\left[\frac{g^2}{\Box}\Psibar
        \frac{i\slask D}{\Box}\Psi\right]
\ee
that can go like $T^2$. The remaining factors
of $1/\Box$ are expanded in powers of $g$ and with the gauge
in \eq{gcond} only the
leading $1/\pa^2$ remain in the high $T$ limit.
The two poles correspond to forward scattering of gauge bosons and fermions,
respectively.
After performing the thermal trace over Lorentz indices we obtain
\be{fermhtl}
        i\Gamma[\Psibar,\Psi]= \frac{g^2C_fT^2}{8}
        \int \frac {d^2\qh} {4\pi}\Psibar
        \,\frac{\gamma_\mu\Qh^\mu}{\Qh\cdot\pa}\,\Psi~~.
\ee
Equation (\ref{fermhtl}), just like \eq{sa}, is gauge dependent and
valid only in the gauge \eq{gcond}. It is, however, straightforward to write
them in a explicitly gauge invariant by expressing $A$ and $\Qh\cdot\pa$
in terms of $F$ and $\Qh\cdot D$ and thereby recovering the standard HTL
effective action.

%
%
%

%
\end{document}